\documentclass[twocolumn,noshowpacs,preprintnumbers,
amsmath,amssymb]{revtex4}
\usepackage{amsfonts}
\usepackage{amssymb}

\newcommand{\proofend}{\raisebox{1.3mm}{%
\fbox{\begin{minipage}[b][0cm][b]{0cm}\end{minipage}}}}

\def\2{{\textstyle\frac{1}{2}}}
\def\ba{\begin{eqnarray}}
\def\ea{\end{eqnarray}}
\def\be{\begin{equation}}
\def\ee{\end{equation}}

\newcommand{\CX}{{\cal{X}}}

\newcommand{\CF}{{\cal{F}}}
\let\a=\alpha   \let\d=\delta
\let\e=\varepsilon   

  \let\l=\lambda \let\m=\mu
   \let\r=\rho

\let\O=\Omega \let\S=\Sigma 
 \let\L=\Lambda \let \G=\Gamma

\def\({\left(} \def\){\right)} \def\<{\langle} \def\>{\rangle}

\newcommand{\md}{\mathrm{d}}

\newcommand{\Lie}{{\mathfrak g}}
\newcommand{\rk}{\mathrm{rank}}

\def\Emd{\!\,{}^E\!\md}
\def\bEmd{\!\,{}^{\bar E}\!\md}
\def\ET{\,\!{}^E\!T}
\def\EL{\,\!{}^E\!L}
\def\bEL{\,\!{}^{\bar E}\!L}

\def\Eg{\,\!{}^E\!g}

\def\O{\Omega}
\def\OE{\Omega_E^\cdot(M)}

\newcommand{\w}{{\wedge}}

\makeatletter
\def\barb{\protect\@barb}
\def\@barb{%
\relax
\bgroup
\def\@tempa{\hbox{\raise.73\ht0
\hbox to0pt{\kern-.1\wd0\vrule width.7\wd0
height.1pt depth.0pt\hss}\box0}}%
\mathchoice{\setbox0\hbox{$\displaystyle\mathrm{b}$}\@tempa}%
{\setbox0\hbox{$\textstyle \mathrm{b}$}\@tempa}%
{\setbox0\hbox{$\scriptstyle \mathrm{b}$}\@tempa}%
{\setbox0\hbox{$\scriptscriptstyle \mathrm{b}$}\@tempa}%
\egroup
}
\makeatother

\newcommand{\wc}{\stackrel{\wedge}{,}}

\usepackage{graphicx}% Include figure files
\usepackage{dcolumn}% Align table columns on decimal point
\usepackage{bm}% bold math

\begin{document}

%\preprint{APS/123-QED}

\title{Algebroid Yang-Mills Theories}
% Force line breaks with \\

\author{Thomas Strobl}
\altaffiliation[]{ITP FSU Jena, Max-Wienpl.~1, 07743 Jena, Germany}
 \email{pth@tpi.uni-jena.de}

%\date
\date{June 23, 2004}% It is always \today, today,

\begin{abstract}
A framework for constructing new kinds of gauge theories is suggested. 
Essentially it consists in replacing Lie algebras by Lie or Courant 
algebroids. Besides presenting %!change
novel topological theories defined in arbitrary 
spacetime dimensions, 
we show that equipping Lie algebroids $E$ with a fiber metric having 
sufficiently many $E$-Killing vectors leads to an astonishingly mild 
deformation of ordinary Yang-Mills theories: Additional fields turn 
out to carry no propagating modes. Instead they serve as moduli parameters 
gluing together in part different Yang-Mills theories. This leads 
to a symmetry enhancement at critical points of these fields, as is also 
typical for String effective field theories.
\end{abstract}

\maketitle

Yang-Mills (YM) theories and Lie group symmetries are part and
parcel of present day fundamental physics. Both of these concepts
are %, moreover,
%very robust against deformations %
``nondeformable''
under very mild assumptions
 %(cf., e.g.,
\cite{defoRep}. %).
 The advent of supersymmetry, possible after
 %slightly
changing the perspective on symmetries, is an example of the
fruitfulness of enlarging that framework. %!change
%that slightly changing the
%perspective may permit a rich enlargement of the framework. 
%the framework
%can  be enlarged profitably if one changes the perspective slightly. 
%The advent of supersymmetry, possible after
%slightly changing the perspective on symmetries, is an example of
%the fruitfulness
%of enlarging that framework (be it observed 
%in experiment or not).
In the present Letter we suggest a possibly similar
broadening, which in its essence replaces Lie groups by Lie groupoids
in the context of Yang-Mills theories; for 
trivial bundles this reduces to replacing the structural Lie algebra
of a YM-theory by a Lie algebroid. In part this generalization
is related to rather old attempts \cite{nonlinYM}
for constructing so-called ``non-linear gauge theories''; the recent
mathematical understanding of Lie algebroids and groupoids
\cite{SilvaWeinstein,Moerd}, however, provides new tools and a new focus
for approaching such a generalization.  

In the theories under discussion generically one encounters 
%Despite the generic appearance of
structure functions in the symmetry algebra, typical
%in particular
for gravitational theories; %in the case under discussion
but %in our case
there will exist a finite dimensional object underlying the
%!change: ``object'' instead of ``concept'' and ``otherwise'' deleted 
infinite dimensional space of symmetries: infinitesimally the symmetries
are generated by sections in a Lie algebroid.
%, a generalization of a Lie algebra recalled  below
Two spacetime dimensions already provides an example
where these concepts have been realized successfully
in terms of Poisson Sigma Models (PSMs) \cite{PSM1,Ikeda}, which, on
the physical side, permit to unify gravitational and YM gauge
theories \cite{TK1}. Using the PSM, as well as %its relation to the
Chern-Simons (CS) theory defined in $d=3$, as a guideline, we will
leave behind low dimensions in %the end of
this Letter and, besides suggesting possibly also
interesting topological models% in higher dimensions
,  permit theories with propagating degrees of freedom. 

We briefly recall some mathematical background \footnote{For further
details cf.~\cite{CMP-BKS,CMP1} or the monographs
\cite{SilvaWeinstein,Moerd}.}: A Lie algebroid 
consists of a vector bundle $E\to M$, a bundle map $\rho \colon
E \to TM$, and a Lie algebra bracket $[\cdot,\cdot]$ on $\G(E)$
satisfying the Leibniz rule $[ s_1, f s_2] = f [ s_1,  s_2] +
(\rho(s_1)f) s_2$. For $M$ a point this reduces to an ordinary Lie
algebra $\Lie$, for $\rho \equiv 0$ to a bundle of Lie algebras, $M$ 
then being a
%!change
parameter space of Lie algebra
deformations. Given a Poisson manifold  $(M,\Pi)$, $\Pi^{ij} \equiv
\{X^i,X^j\}$, one obtains a less obvious Lie algebroid by means of
$E := T^*M$, $\rho(\a) := \a_i \Pi^{ij} \partial_j$, and
 $[\md f, \md g ] := \md \{f , g\}$. 

For later use we remark that 
the image of $\rho$ is integrable so that $M$ is foliated into orbits. 
Moreover, due to the Leibniz rule, the bracket 
%between sections 
reduces to a fiberwise
Lie algebra structure for elements in the kernel of $\rho$,
which is isomorphic for any two points in the same orbit.
%and for any two  
%points in the same orbit the algebras are isomorphic. 

%$X,X' \in M$ in the same orbit
% $\ker \rho \subset E$ carries a Lie algebra  
 
In local coordinates %$X^i$, $i=1,\ldots \dim{M}$
$(X^i)_{i=1}^{\dim{M}}$ and frame $(b_I)_{I=1}^{\rk{E}}$ the
Lie algebroid data %can be
are encoded in %terms of
structural functions
$\rho^i_I(X)$, %and
$C^I_{JK}(X)$, where $\rho(b_I) \equiv \rho^i_I \partial_i$,
$[b_I,b_J] \equiv C^K_{IJ} b_K$. %, satisfying a particular set of
%differential compatibility conditions (cf., e.g., \cite{CMP1}).
%It is only a particular feature of the Poisson Lie algebroid
%that the anchor $\rho$ determines 
%In the Poisson case, $b_I \sim \md X^i$, $\rho^{j}_I \sim \P^{ij}$,
% and $C_{JK}^I \sim \P^{jk}{}_{,i}$, so that the latter functions are
% determined completely by the anchor $\rho$.

Guided by the other obvious example of a Lie algebroid, $E=TM$ with
$\rho = \mathrm{id}$, one may introduce differential geometrical
notions on Lie algebroids. With $b^I$ denoting the dual basis in
$E^*$, the Leibniz extension of $\Emd X^i = b^I \rho^{i}_I (X)$ and
$\Emd b^I = -\frac{1}{2} C_{JK}^I (X) b^J \wedge b^K$ defines a
generalization of the de Rham differential on the space of $E$-forms
$\Gamma(\Lambda^\cdot E^*) \equiv \OE$; $\Emd^2=0$ entails all the 
differential compatibility conditions to be satisfied by the
structural functions introduced above. Also one may be interested in
differentiation along sections of $E$: On $E$-tensors this
may be done by means of a generalized $E$-Lie derivative. For $s_1,s_2
\in \Gamma(E)$ e.g.~$\EL_{s_1} s_2 := [s_1,s_2]$ while on $\OE$ one
uses $\EL_s = \Emd \; \iota_s + \iota_s \, \Emd$. For sections $\psi
\in \Gamma(V)$, $V$ being any other vector bundle over $M$, one
introduces a ``contravariant'' derivative
$^{E}\nabla \colon \Gamma(V) \to \Gamma(E^*) \otimes \Gamma(V)$ by
a straightforward extrapolation of the axioms of a standard covariant
derivative.

Besides a Lie algebra, the definition of a YM action requires a
non-degenerate ad-invariant scalar product. The obvious generalization
would be a fiber metric $\Eg$ %satisfying
with
$\EL_s \Eg =0$, $\forall s \in
\Gamma(E)$. However, for $\rho \not \equiv 0$ this requires
%is possible only for
Courant algebroids (CAs): As before, $\rho \colon E \to TM$ and there is a
bracket for %on the
sections %of $E$
satisfying the Leibniz property
w.r.t.~itself as well as multiplication by $C^\infty(M)$. %functions. 
% and w.r.t.~itself.
In addition
there is $\Eg$ with
%such that
$\rho(s_1)\, \Eg(s_2,s_3) =  \Eg([s_1,s_2],s_3)+
\Eg(s_2,[s_1,s_3])$, $\forall s_i \in
\Gamma(E)$. If the bracket were required to be antisymmetric,
$E$ would become a Lie algebroid and, as one may check, 
%!change
%mentioned above,
the existence of  $\Eg$ would require
$\rho\equiv 0$. The final axiom of a CA, circumventing this rather
trivial case, is $\Eg([s,s],s') := \rho(s') \, %\left(
\Eg(s,s)
%\right)
$.
Although it will be possible to define a Courant algebroid YM (CAYM)
theory, and we will do so at the end, % of this letter,
we show that
%using
the better understood Lie algebroids
are
sufficient for our purposes. 

To develop a  framework for the fields of a YM-type theory %under
%discussion
we first draw on the relation between
%!change
 the PSM and the CS gauge
theory. We argue that, viewed from the correct perspective,
they are essentially the same; one needs only to exchange the target Lie
algebroid $E$, which is $T^*M$, $M$ Poisson, in the PSM and
$\Lie$ in the CS-theory \footnote{Further
details
%!change
on this and the following two paragraphs
can be found in %For details cf.~
\cite{CMP-BKS}.}.
Let $\S$ denote the spacetime under consideration. Then in
both cases the fields of the theory are vector bundle morphisms $a
\colon T \S \to E$ \footnote{In the present paper we only consider
trivial bundels. The case of non-trivial bundles will be discussed
elsewhere \cite{AT}.}. Such maps are specified by a base map $\CX \colon
\S \to M$ and a section $A \in \O^1(\S,\CX^*E)$. In local terms,
$\CX$ corresponds to scalar fields $X^i(x)$ and $A = A^I \otimes
\barb_I$, where $(\barb_I)$ denotes the basis in $\CX^*E$ induced by
$(b_I)$ in $E$ and $A^I$ is a collection of 1-forms on $\S$.  
In the Poisson case, $b_I \sim \md X^i$, and we  recover the
fundamental fields $(X^i,A_i)$ used to define the PSM. 
%theory.
In the CS
theory %, on the other hand,
$\CX$
%maps all of $\S$ to a single point
contains no information, mapping all
of $\S$ to a single point, and  $\CX^*E \cong E = \Lie$, so that $A$
becomes the Lie-algebra valued connection 1-form.

Next we turn to the field equations. The transpose of a vector bundle
morphism $a$ is a map between sections: $a^T \colon \OE \to
\O^\cdot(\S)$. E.g.~$a^T(X^i) = \CX^* X^i \cong X^i(x)$, $a^T(b^I)=A^I$.
The vector bundle morphism
$a$ is also a morphism of Lie
algebroids, iff the operator 
$\CF := \md \, a^T - a^T \, \Emd$ %\colon  \OE \to \O^\cdot(\S)$
vanishes identically on $\OE$. Specializing 
%It is now straightforward to check
%that in both cases under considerations the field equations reduce to
%$\CF=0$. Indeed, specializing 
\ba F^i := \CF(X^i) = \md X^i - \rho^i_I A^I, \label{Fupper}
\\ \label{Flower}
F^I :=  \CF(b^I) =
\md A^I + \frac{1}{2} C^I_{JK} A^J \wedge A^K 
\ea
%apply $\CF$ to local coordinates $X^i$ and a local
%frame $b^I$,
%\be \CF
to
%both cases,
$E \cong \Lie$ and $E \cong T^*M$ 
(with
%, (in the second case
$\rho^j_I \sim
\Pi^{ij}$, % and 
$C^I_{JK} \sim \Pi^{jk}{}_{,i}$), one recovers the
respective 
field equations from $\CF=0$. 
%regains the respective field equations

Finally, two solutions $a$, $a'$ to $\CF = 0$ are gauge equivalent,
iff they are homotopic, $a \sim a'$. Infinitesimally this implies 
%\ba
%\d_\e X^i = \rho^i_I \e^I \, ,\label{Sym01}\\
%\d_\e A^I \approx  \md\e^I + C^{I}_{JK} A^J \e^K \,\label{Sym02} ; 
%\ea
$\d_\e X^i = \rho^i_I \e^I$, $\d_\e A^I \approx  \md\e^I + C^{I}_{JK}
A^J \e^K=: \delta^{(0)}_\e A^I$ with $\e \in \O^0(\S,\CX^*E)$, where $\approx$ is
chosen so as to stress 
the on-shell character of the equation
%that one has an on-shell equation only
(resulting from an on-shell concept). This is readily
recognized as the gauge symmetries of the PSM and CS %theory
upon specialization. 

We now  address the question whether one can find a
topological theory for any dimension $d$ of $\S$ and 
%in particular also for 
any choice of the target Lie algebroid $E$ such that
the morphism property $\CF=0$ is contained in the field equations and
the homotopy in the gauge symmetries.
%the field
%equations contain the morphism property $\CF=0$ and the gauge symmetry
%includes the above specified homotopy.
%?The answer is in the affirmative:
Introducing auxiliary $(d-1)$-forms $B_i$ and $(d-2)$-forms $B_I$,
\be S_{\mathrm{LABF}} := \int_\S B_i \wedge F^i + B_I \wedge F^I \, , 
\label{BF} \ee
obviously leads to the desired field equations and---no
more so 
%that 
trivial, relying heavily on $\Emd^2=0$---it
%the differential identities
%satisfied by the structural functions of $E$
 is
invariant with respect to the above gauge symmetries
% when complemented by
for
$\d_\e B_I := C^J_{KI} B_J  \e^K$ and
$\d_\e B_i := -\e^I (\rho^j_{I,i} B_j + C^K_{IJ,i} B_K \wedge A^J)$.

For $E = \Lie$ this action  reduces to that of a
non-abelian BF-theory, for $E=T^*M$, $M$ Poisson, it was
suggested already in \cite{Izawa}. It is topological due to further
gauge symmetries on the $B$-fields, following from Bianchi type
identities for the ``curvatures'' $F^i$ and $F_I$. They
% latter ones
can be obtained
%either by using the structural equations of $E$
%generalizing the Jacobi identity for a Lie algebra $\Lie$, or, much
%simpler,
most easily by applying the obvious relation $\md \circ  \CF = - \CF
\circ \Emd$
%(it is here where $\Emd^2=0$ is used)
%(following from $\md^2=0=\Emd^2$) 
to $X^i$ and $b^I$, respectively. In
the first case this gives $\md F^i \equiv - \rho^i_I F^I + \rho^i_{I,j}
A^I \wedge F^j$, leading to the independent gauge symmetry %: (besides
 ($\d_\l X^i=0 =\d_\l A^I$):% ) % for $TM$.}
\be \label{lambdasymm}
%\begin{eqnarray}
\d_\l B_i = \md \l_i + \r^j_{I,i} A^I \w \l_j \, ,  \;\, \d_\l B_I =
(-1)^{d+1} \rho^i_I \l_i  \, . \ee % \end{eqnarray} 
Likewise, %due to
$\md F^I  + C^I_{JK} A^J \wedge F^K \equiv
\frac{1}{2} C^I_{JK,i} A^J \wedge A^K \wedge F^i $ gives
%leads to
%, the transformations
$\delta_\mu B_I = \md \mu_I - C^J_{KI} A^K \wedge
\mu_J$, $\delta_\m B_i = \frac{1}{2}C^I_{JK,i} A^J \wedge \m_I \wedge A^K$,
as an invariance of  (\ref{BF}) for any
$\mu \in \O^{d-3}(\S,\CX^* E^*)$.
%leave  (\ref{BF}) invariant for any
%$\mu = \mu_I \otimes \barb^I \in \O^{d-3}(\S,\CX^* E^*)$.
 
%In the second case, this leads to
%$\md F^I =  -C^I_{JK} A^J \wedge F^K +
%\frac{1}{2} C^I_{JK,i} A^J \wedge A^K \wedge F^i $. Correspondingly,
%there exists a gauge symmetry, independent from the above one, of the
%form $\delta_\mu B_I = \md \mu_I + C^J_{KI} A^K \wedge
%\mu_J$, $\delta_\l B_i = -C^I_{JK,i} A^J \wedge \l_I \wedge A^K$,
%with $\mu = \mu_I \otimes \barb^I \in \O^{d-3}(\S,\CX^* E)$. 
%Likewise $\md F^i = - \rho^i_I F^I + \rho^i_{I,j} A^I \wedge F^j$
%gives the symmetry:

To obtain a general framework for constructing action functionals for
a field $a \colon T\S \to E$, reducing to ordinary YM-theory for
$E=\Lie$, we need to address the meaning of %the left-hand side of
(\ref{Fupper}) and (\ref{Flower}) as well as what the off-shell gauge
transformations are \footnote{A detailed account of these issues is given in 
\cite{CMP-BKS,AT}.}. 
While $F^i$ and $\d_\e X^i$ have a well-defined
meaning, e.g.~$\Phi := F^i \otimes \partial_i \in \O^1(\S,\CX^* TM)$
corresponding to the bundle map $\phi = \CX_* - \rho \circ a
%$ from  $T\S$ to $TM$
\colon T\S \to TM$, this is the case for neither $F^I$ nor
$\delta^{(0)}_\e A^I$;
%the $A$-symmetries;
despite their above naive use 
%, as written,
both of them are frame-dependent.% \cite{CMP-BKS}. 

This can be cured by introducing a connection $\nabla$ on $E$. %\to M$. 
The combination $F^I + \G^I_{iJ} F^i \w A^J=: F_{(\G)}^I$, where
$\nabla b_I \equiv \G^J_{iI} \, \md X^i \otimes b_J$, transforms
correctly under a change of frame. In explicitly covariant terms this
yields 
\be  F_{(\G)}^I = (D A)^I - \frac{1}{2} \ET^I_{JK} A^J \w A^K \, ,
\ee
where $D$ is the canonical exterior
covariant derivative on $\O^\cdot(\S,\CX^* E)$ induced by $\nabla$ and
$\ET$ is the $E$-torsion of  $ ^{E}{}\nabla_s :=
\nabla_{\rho(s)}$. Now $F=F_{(\G)}^I \otimes \barb_I \in
\O^2(\S,\CX^*E)$, a 2-form on spacetime with values in what replaces
the Lie algebra, also comes from a bundle map $f \colon \L^2 T\S \to
E$.%\cite{AT-curv} 

Covariant  $A$-gauge symmetries %must
have the form
$\d_\e A^I =  \d_\e^{(0)} A^I + \l^I_i F^i$ 
%,$\d_\e^{(0)} A^I \equiv  \md\e^I + C^{I}_{JK}
%A^J \e^K$,
for some $\e$-dependent
%choice of
$\l^I_i$. One option is to use the above (or any other)
connection, 
setting $\l^I_i = \G^I_{iJ} \e^J$. Another, qualitatively different
possibility is   $\l^I_i = - \e^I_{,i}$, where $\e^I$ now is
viewed as pullback of $\e^I(x,X)$ by $\CX$ 
\footnote{One may  
%The two alternatives may be
compare to the two 
qualitatively different lifts of tangent vectors from some manifold $N$ to 
$TN$ as a covariant or as a Lie derivative; the first option requires the 
additional structure of a connection on $N$, the second one is canonical 
provided 
the given vector is extended infinitesimally to a vector field. In our case 
$\{ \CX \colon \S\to M\}$ 
plays the role of $N$, the given tangent vector is 
$\d_\e X^i = \rho_I^i \e^I$, and $\{ a \colon T\S\to E\}$
replaces $TN$. Although these map spaces are infinite dimensional, our lifts 
are goverened by finite dimensional lifts via appropriate
 left actions.  }. 
This has a nice geometrical interpretation: Replacing $E$ 
by the exterior sum Lie algebroid $\bar E=T\S \boxplus E$ over $\S
\times M$,   $\bEmd = \md +
\Emd$, 
and using the graph $\bar a = \mathrm{id} \boxplus a$, the gauge 
symmetries are generated by sections $\e$ of $\bar E$ with values in  
$E$-fibers by means of a left action of $\bar E$
%!change
on itself: 
$\delta^{(2)}_\e \bar a^T =\bar a^T \circ \bEL_\e$.  Locally we can 
always choose a flat connection $\G$ or an $X$-independent prolongation of 
$\e \in \O^0(\S,\CX^*E)$; then in both cases one reobtains  
$\d_\e^{(0)} A^I$. But %However,
for  non-flat bundles $E$,
%this cannot be done 
this is not possible
globally.

%The second option does not require an additional structure such as a
%connection and
%%require a connection and, due to the appearance of an $E$-Lie
%%derivative,
%has  computational advantages; for the price that, in contrast to 
%the first one, it depends on the extension of 
%$\e^I(x)$ away from $X(x)$. One may  
%%The two alternatives may be
%compare to the two 
%qualitatively different lifts of tangent vectors from some manifold $N$ to 
%$TN$ as a covariant or as a Lie derivative; the first option requires the 
%additional structure of a connection on $N$, the second one is canonical 
%provided 
%the given vector is extended infinitesimally to a vector field. In our case 
%$\{ \CX \colon \S\to M\}$ 
%plays the role of $N$, the given tangent vector is 
%$\d_\e X^i = \rho_I^i \e^I$, and $\{ a \colon T\S\to E\}$
%replaces $TN$. Although these map spaces are infinite dimensional, our lifts 
%are goverened by finite dimensional lifts via appropriate
% left actions. 

Now we are able %in the position
to obtain explicitly 
covariant, %and
globally defined action functionals. We 
first
reconsider %the
topological theories from this enhanced 
perspective. The covariant analogue
of (\ref{BF}) would be $\int_\S 
b_i \wedge F^i + B_I \w F^I_{(\G)}$, where $b \in \O^{d-1}(\S,\CX^*T^*M)$, 
$B \in \O^{d-2}(\S,\CX^*E^*)$. But whatever the choice for $\G$ is, flat or 
non-flat, in any local patch the redefinition $B_i := b_i + \G^I_{iJ} B_I \w 
A^J$ makes the connection disappear, bringing the action into the form 
(\ref{BF}). The globally well-defined form of the
transformations of the Lagrange multiplier fields also depend 
on  $\l^I_i$. We display one of them,
$\delta B_I = - C^K_{IJ} B_K \e^J - \rho^i_I B_J
\l^J_i$, as it is needed also in the  context of (\ref{LAYM})
below. 
%Since needed in the context of (\ref{LAYM}) below, we
%display one 
%depends on the form of the global off-shell symmetry, 
%parametrized by $\l^I_i$.
%(For checking gauge invariance of
%(\ref{LAYM}) below, we display only
%$\delta B_I = - C^K_{IJ} B_K \e^J - \rho^i_I B_J
%\l^J_i$.)
Retrospectively,  the previous considerations can be justified 
and backed up by an underlying global construction.

%Albeit that $B_i$ are not 
%components of a section in a vector bundle, in contrast to 
%$b=b_i \otimes \barmd X^i$, the patchwise computational 
%simplification in the transformed fields is considerable.

%Let us briefly return to the analogy between CS and the PSM. It goes even 
%further: 

For particular choices of $E$ there may exist %a dimension of $\S$ such that 
topological actions 
 producing Lie algebroid morphisms 
from $T\S$ to $E$ up to homotopy without any auxiliary fields.
But %the dimension
$d=\dim \S$ %of  $\S$
will depend on the choice of $E$: For $E$ a quadratic Lie algebra 
$\Lie$, one has the CS theory in $d=3$, for $E=T^*M$, $M$ Poisson, one has 
the PSM in $d=2$. The analogy between the two models goes even further: 
The CS theory can be regarded as being induced 
on the  boundary $\S = \partial \widetilde \S$ of 
a spacetime $\widetilde \S$  with an extra dimension, using an 
%stemming from an 
$F \wedge F$-action;
%on a spacetime $\widetilde \S$ 
%with an extra dimension, as the induced
%action on the respective boundary $\S = \partial \widetilde \S$; 
here an ad-invariant metric on $E=\Lie$ is needed to 
%This corresponds to $E=\Lie$, and in the action 
%an adinvariant metric is used to 
contract the Lie algebra indices of $F=F^I \otimes \barb_I$. 
On the other 
hand, for $E=T^*M$, no metric is needed to contract $F^i$ with $F_{i(\G)}$. 
This now
gives a 3-form,
%the connection
 $\G$ drops out if $\nabla$ is chosen
 torsionfree, 
and indeed % one obtains
\be S_{\mathrm{PSM}}= \int_{\widetilde \S} F^i \w F_i \,  \label{PSM}\ee
induces the Poisson Sigma Model %induced
 on the  two-dimensional boundary  $\S = \partial \widetilde \S$. 
Whereas  the choice $E=\Lie$ requires $d=3$ for a CS-theory,
for $E=T^*M$, $M$ Poisson, % such a theory,
a ``Poisson-CS-theory'' naturally lives in two dimensions
and coincides with the PSM.
%the PSM may be viewed as a CS-type-theory, 
%just that %for
%$E=\Lie$ requires $d=3$
%such a theory lives in three dimensions
%while  for $E=T^*M$, $M$ Poisson, such a theory
%naturally lives in two dimensions. 
We note %in parenthesis
as an aside that similarly to the Pontrijagin class 
$\langle F 
\stackrel{\wedge}{,} F\rangle$, also $F_i \w F^i$ corresponds
to a charateristic class,
%\emph{CHECK} 
one associated with 
Poisson fibrations \cite{AT}. %\emph{CHECK}
%e will come back to this
%elsewhere

If one does not permit WZ-contributions to the action \cite{Ctirad}
%and does not permit
nor auxiliary metrics, 
the PSM is the universal purely bosonic 
topological theory in two dimensions \cite{Izawa}.  
%at least if one. 
Correspondingly, in $d=2$ the 
LABF-theory (\ref{BF}) must 
%\emph{must} 
be a particular PSM. Indeed, generalizing 
the old observation that ordinary BF-theories are PSMs for $M=\Lie^*$, in two 
dimensions one recovers the LABF-theory from the PSM for the particular 
choice $M=E^*$. The dual of a Lie algebroid is canonically a Poisson manifold, 
locally $\{b_I,b_J\}=C_{IJ}^K b_K$, $\{X^i ,X^j \} =0$, and 
$\{b_I ,X^i\} = \rho^i_I $. $A^I$ and $B_i$ collect into the 1-form fields 
of the PSM,  $X^i$, $B_I$ into the scalar ones, and also the gauge symmetries 
(\ref{lambdasymm}) are included in this description.  %ueberprueft.

There exists a likewise universal \cite{IkedaCSM} topological action
in three dimensions, reducing to the CS-theory for $M$ a point, 
which one might call the Courant Sigma Model
(CSM). Using the description \cite{DimaCourant} of Courant
algebroids as particular QP-structures, the action can be obtained
most easily by the AKSZ-approach \cite{AKSZ}: \footnote{Adapted
from: D.~Roytenberg, talk FSU Jena, May 2005.}
\be S_{\mathrm{CSM}}=\int_\S B_i \w F^i + \frac{1}{2}
 \left\langle A \stackrel{\w}{,} 
\md A + 
\frac{1}{3} [A \stackrel{\w}{,} A] \right\rangle \, ,
\label{CSM}
\ee
%!ACHTUNG (NUN wieder ueberfluessig; nur wenn der Faktor 1/2
%weggegeben wird): wegen des fehlenden Faktors, muss man nun bei der
%Identifikation unten die LABF-Theorie durch 2 dividieren und
%das B-Feld dort mit 2$B_i$ hier identifizieren. 
where $[ \cdot , \cdot]$ and  $\langle \cdot , \cdot\rangle$ denote bracket 
and fiber metric in a Courant algebroid, respectively, 
%$A \in \O^1(\S,\CX^*E)$,
and $F^i$ agrees with (\ref{Fupper}). 
%
%Although in the construction 
%of  (\ref{CSM}) an orthonormal frame was used, and as 
%written the action is not 
%in an explicitly covariant form, 
%
%The action (\ref{CSM}) is not covariant;  \emph{CHECK: Probably wrong!} 
%the CSM is of this form only when $b_Z$ in $A=A^Z \barb_Z$ is an 
%orthonormal frame of  $\langle \cdot , \cdot\rangle$ and 
%$\md A=\left(\md A^Z \right) \barb_Z$. 
The LABF-theory in $d=3$ is a particular CSM, 
where the Courant algebroid is $E \oplus E^*$ with anchor $\rho \oplus 0$, 
canonical fiber metric, % the canonical one,
and bracket $[s \oplus u , s' \oplus u']
= [s,s'] \oplus \left(\EL_s u' - \iota_{s'} \, \Emd u \right)$; 
$A^I b_I \oplus B_I b^I$ combine into the $A$-field of ({\ref{CSM}).
%,and $B_i$ agrees with $A_i$. 
%\emph{CHECK if symmetries agree.} 

We now turn to nontopological Lie algebroid 
theories, generalizing standard 
YM-gauge theories.

Our main proposal in this context is the action
\be S_{\mathrm{LAYM}} = \int_\S B_i \w F^i - \frac{1}{2} \left\langle F \wc * F
\right\rangle
%\frac{1}{2} \Eg_{IJ} F^I 
%\wedge * F^J
\, .
\label{LAYM}
\ee
Here the fiber metric $\langle \cdot , \cdot \rangle \equiv
\Eg(\cdot,\cdot)$
is assumed to admit %!change
%possessy
%$\rk E$ independent 
a (possibly overcomplete) basis of sections $\psi_A$ satisfying 
$\EL_{\psi_A} \Eg=0$. %, $A=1,\ldots,\rk E$.
Then %the action 
(\ref{LAYM}) is invariant under gauge transformations (of the Lie type) 
with respect to $\e^I=\e^A(x) \psi^I_A(X(x))$ for arbitrary $\e^A(x)$; 
the symmetries are a module of $C^\infty(\S)$ \footnote{Tensored by functions
on $M$ constant along the orbits.}, %or  $C^\infty(\S)
%\otimes C^\infty()$, where
but no more 
of all of $C^\infty(\S\times M)$ as in the topological theories. 
(This reminds one of a similar relation between Kac Moody algebras and 
current algebras governing topological models \cite{AS}.)

Due to the presence 
of the first term, we are permitted to drop any explicit
$\G$-dependence in (\ref{LAYM}). 
%We were permitted to drop any $\G$-dependence due to the presence 
%of the first term.
Indeed, (\ref{LAYM}) results from the explicitly 
covariant LABF-theory upon addition of a term like
$\Eg^{IJ} B_I \w * B_J$,
shifting $b_i$ before eliminating $B_I$. Likewise, 
in the 
construction of (\ref{CSM}) an orthonormal frame was used, and, as written, 
it is not explictely covariant, but again %also there
the presence 
of the first term makes it invariant w.r.t.~%with respect to %.r.t.~any 
a change of frame. %last statement}
Covariance is a very useful tool in the discussion of  (\ref{LAYM}), 
already when checking the above mentioned gauge invariance. Also it permits 
 to put $\Eg_{IJ}$ constant or to bring the Lie algebroid structural functions 
into some particularly simple %local
form. 

Somewhat miraculously, contrary to first appearance, the action  (\ref{LAYM}) 
contains no other propagating degrees of freedom as those present in 
\emph{ordinary} YM gauge theories. This may be seen as follows: The 
$B_i$-field equation $F^i\equiv\md X^i - \rho^i_I A^I
=0$ restricts the image of 
the map $\CX\colon \S \to M$ to lie in an orbit of $E$ in $M$. Then one may 
use the gauge symmetries to reduce $\CX$ %!change
to the respective homotopy classes $[\CX]$. 
In the trivial class we may put $X(x)=X_0=const$. (Nontrivial classes will 
correspond to a new kind of YM instanton sectors.) Returning to $F^i=0$, 
%(\ref{Fupper}),
$A$ now is forced to lie in the kernel of $\rho$ at 
$X_0 \in M$. On the other hand, variation of $S_{\mathrm{LAYM}}$
along directions of 
$A$ in $\ker \rho$ contain no $B$-field and reduce to the ordinary YM-field 
equations for $\Lie = \ker \rho|_{X_0}$. And %Moreover, 
the residual $\e$-gauge symmetry
is readily recognized as the usual nonabelian one of $\Lie$. 

It remains to show that $B_i$ contains no propagating modes. This is somewhat 
more subtle and  will be detailed elsewhere. (It may be illustrative 
for the reader, however, to check the particularly simple cases $E=TM$ and 
$E$ a bundle of Lie algebras.) Essentially it works as follows: 
the $A$-variation determines $B_i$ up to parts in the conormal bundle of the 
respective orbit (in terms of $X_0$ and the $\Lie$-connection $A$). 
On the other
hand, the variation w.r.t.~$X$ yields a field equation containing $\md B_i$. 
To kill the remaining parts of $B$, up to possibly global modes when 
$\S$ has nontrivial topology, one needs an additional symmetry. This is 
provided by  a relict of (\ref{lambdasymm}): 
(\ref{LAYM}) is invariant under a $B_i$ transformation for 
any $\l$ taking values in the conormal bundle of an orbit. 

We have $\dim \Lie = \rk E - \dim {\cal O}$, where ${\cal O}$ 
denotes the 
orbit through $X_0$. Correspondingly, at singular leaves one obtains 
a higher dimensional structural group of the respective YM-theory in 
comparison with points $X_0$ in nearby leaves. At such exceptional orbits 
(they are of measure zero in $M$) it is also not clear if one should 
eliminate all of the $B_i$-modes as mentioned above, since this in part can 
correspond to $\lambda$'s %!change
that are distributions on $M$.
%\emph{??CHECK}
Classically there 
is no interaction between the LAYM-theories defined over different leaves 
of the foliation. On the quantum level the situation may be more intricate 
for what concerns singular leaves of the foliation, as we see in 
the Kontsevich 
formula \cite{Kontsevich}. In fact, in $d=2$ the action (\ref{LAYM}) becomes a 
particular almost topological PSM %, as proposed also in
\cite{PSM1}, resulting 
from  the topological one by the addition of a Casimir function multiplied by 
a volume form on $\S$.

The above strategy used to obtain LAYM-theories may be extended to 
Courant algebroids.
Instead of the PSM one then starts with the CSM 
(\ref{CSM}), and, as before, uses its symmetries and the 
l.h.s.~of its field equations in arbitrary dimensions. % of $\S$. 
In this way one obtains a CABF-theory, which becomes a CAYM-theory 
by adding quadratic terms in the Lagrange multipliers. % fields.
If one
uses only the multiplier for the 2-form field strength, the given scalar
product is sufficient for gauge invariance. % without any restriction.
Again one obtains YM-theories, but now also gauge
theories for 2-form gauge fields can be constructed. This is
particularly interesting in view of their presence in String effective
field theories, but also their and CAs believed relation
to gerbes. Higher degree gauge fields will
arise when climbing up in
the ladder of dimensions, at the price of more complicated algebroids. 

%more complicated algebroids will be induced and higher degree
%forms take part in a more general notion of connection. 

%a term of the form $\langle B \wc *B \rangle$, which is gauge 
%invariant  without any further restriction on the fiber metric. 

% Note that 
%in this context one uses also 2-forms for what replaces the connection, which 
%became interesting from several perspective lately. Climbing up in the ladder
%of dimensions more complicated algebroids will be induced and higher degree
%forms take part in a more general notion of connection. 

In the present Letter we showed,  
however, that in all dimensions we can fruitfully restrict to Lie 
algebroids. The nonexistence of invariant metrics for nontrivial $\rho$ 
can be circumvented by the weaker condition for the existence 
of $E$-Killing vectors; also, in the context of gravitational
theories, a likewise problem was circumvented by introducing an
$E$-Riemannian foliation on the target \cite{CMP1}. Lie algebroids are well 
understood \cite{SilvaWeinstein,Moerd}, 
recently even the necessary and sufficient conditions for integrating 
them to Lie groupoids have been clarified \cite{CFm1}, so that  
%With such tools at hand
a generalization of e.g.~Wilson loops comes %!change
into reach. 

The above tools permit to construct also other 
Lie algebroid theories. For example, % E.g.,
%at least
 in some cases,
one can add a term quadratic
in $B_i$ to (\ref{LAYM}) not spoiling  gauge invariance.
Physically this corresponds to adding Higgs fields and/or giving mass
to some of the gauge fields. 
%We remark in parenthesis that in $d=2$ such a theory 
%reduces to String Theory (formulated
%as sketched briefly in the introduction to 
%\cite{CMP1}) with degenerate open string metric. 
%But it will be interesting also to further extend 

In two spacetime dimensions the PSM unifies gravitational and 
YM-fields into a common framework. Super-symmetrization is 
obtained by using graded Poisson manifolds as target
\cite{TSsugra}. It remains
to be seen in how far 
this picture can be extended to higher dimensions.

%, or
%whether it can even give a better grasp on some of the String
%dualities.

%would be 
%marvellous if this picture can be extended to higher dimensions, at least 
%in part. 

\begin{acknowledgments}
I am grateful to M.~Bojowald and in particular to A.~Kotov for
collaboration and discussions on the mathematical basis of the 
present considerations. I am also grateful to %M.~Gr\"utzmann,
A.~Alekseev, C.~Mayer and T.~Mohaupt for discussions and
remarks on the manuscript.
%and A.~Wipf for discussions.
\end{acknowledgments}

%\bibliography{bibfile}% Produces the bibliography via BibTeX.

\begin{thebibliography}{19}
\expandafter\ifx\csname natexlab\endcsname\relax\def\natexlab#1{#1}\fi
\expandafter\ifx\csname bibnamefont\endcsname\relax
  \def\bibnamefont#1{#1}\fi
\expandafter\ifx\csname bibfnamefont\endcsname\relax
  \def\bibfnamefont#1{#1}\fi
\expandafter\ifx\csname citenamefont\endcsname\relax
  \def\citenamefont#1{#1}\fi
\expandafter\ifx\csname url\endcsname\relax
  \def\url#1{\texttt{#1}}\fi
\expandafter\ifx\csname urlprefix\endcsname\relax\def\urlprefix{URL }\fi
\providecommand{\bibinfo}[2]{#2}
\providecommand{\eprint}[2][]{\url{#2}}

\bibitem[{\citenamefont{Barnich et~al.}(2000)\citenamefont{Barnich, Brandt, and
  Henneaux}}]{defoRep}
\bibinfo{author}{\bibfnamefont{G.}~\bibnamefont{Barnich}},
  \bibinfo{author}{\bibfnamefont{F.}~\bibnamefont{Brandt}}, \bibnamefont{and}
  \bibinfo{author}{\bibfnamefont{M.}~\bibnamefont{Henneaux}},
  \bibinfo{journal}{Phys. Rept.} \textbf{\bibinfo{volume}{338}},
  \bibinfo{pages}{439} (\bibinfo{year}{2000}).
  %,\eprint{hep-th/0002245}.
  
\bibitem[{\citenamefont{Schoutens et~al.}(1991)\citenamefont{Schoutens, Sevrin,
  and van Nieuwenhuizen}}]{nonlinYM}
\bibinfo{author}{\bibfnamefont{K.}~\bibnamefont{Schoutens}},
  \bibinfo{author}{\bibfnamefont{A.}~\bibnamefont{Sevrin}}, \bibnamefont{and}
  \bibinfo{author}{\bibfnamefont{P.}~\bibnamefont{van Nieuwenhuizen}},
  \bibinfo{journal}{Phys. Lett.} \textbf{\bibinfo{volume}{B255}},
  \bibinfo{pages}{549} (\bibinfo{year}{1991}).

\bibitem[{\citenamefont{da~Silva and Weinstein}(1999)}]{SilvaWeinstein}
\bibinfo{author}{\bibfnamefont{A.~C.} \bibnamefont{da~Silva}} \bibnamefont{and}
  \bibinfo{author}{\bibfnamefont{A.}~\bibnamefont{Weinstein}},
  \emph{\bibinfo{title}{{G}eometric {M}odels for {N}oncommutative {A}lgebras}},
  vol.~\bibinfo{volume}{10} of \emph{\bibinfo{series}{Berkeley Mathematics
  Lecture Notes}} (\bibinfo{publisher}{American Mathematical Society,
  Providence, RI}, \bibinfo{year}{1999}), \bibinfo{note}{available at
  http://www.math.berkeley.edu/~alanw/}.

\bibitem[{\citenamefont{Moerdijk and Mr{\v{c}}un}(2003)}]{Moerd}
\bibinfo{author}{\bibfnamefont{I.}~\bibnamefont{Moerdijk}} \bibnamefont{and}
  \bibinfo{author}{\bibfnamefont{J.}~\bibnamefont{Mr{\v{c}}un}},
  \emph{\bibinfo{title}{Introduction to foliations and {L}ie groupoids}},
  vol.~\bibinfo{volume}{91} of \emph{\bibinfo{series}{Cambridge Studies in
  Advanced Mathematics}} (\bibinfo{publisher}{Cambridge University Press},
  \bibinfo{address}{Cambridge}, \bibinfo{year}{2003}), ISBN
  \bibinfo{isbn}{0-521-83197-0}.

\bibitem[{\citenamefont{Schaller and Strobl}(1994)}]{PSM1}
\bibinfo{author}{\bibfnamefont{P.}~\bibnamefont{Schaller}} \bibnamefont{and}
  \bibinfo{author}{\bibfnamefont{T.}~\bibnamefont{Strobl}},
  \bibinfo{journal}{Mod. Phys. Lett.} \textbf{\bibinfo{volume}{A9}},
  \bibinfo{pages}{3129} (\bibinfo{year}{1994}).
  %, \eprint{arXiv:hep-th/9405110}.

\bibitem[{\citenamefont{Ikeda}(1994)}]{Ikeda}
\bibinfo{author}{\bibfnamefont{N.}~\bibnamefont{Ikeda}}, \bibinfo{journal}{Ann.
  Phys.} \textbf{\bibinfo{volume}{235}}, \bibinfo{pages}{435}
  (\bibinfo{year}{1994}).%, \eprint{arXiv:hep-th/9312059}.

\bibitem[{\citenamefont{Kl{\"o}sch and Strobl}(1996)}]{TK1}
\bibinfo{author}{\bibfnamefont{T.}~\bibnamefont{Kl{\"o}sch}} \bibnamefont{and}
  \bibinfo{author}{\bibfnamefont{T.}~\bibnamefont{Strobl}},
  \bibinfo{journal}{Class. Quant. Grav.} \textbf{\bibinfo{volume}{13}},
  \bibinfo{pages}{965} (\bibinfo{year}{1996}), \bibinfo{note}{erratum ibid. 14
  (1997) 825}.%, \eprint{arXiv:gr-qc/9508020}.

\bibitem[{\citenamefont{Izawa}(2000)}]{Izawa}
\bibinfo{author}{\bibfnamefont{K.~I.} \bibnamefont{Izawa}},
  \bibinfo{journal}{Prog. Theor. Phys.} \textbf{\bibinfo{volume}{103}},
  \bibinfo{pages}{225} (\bibinfo{year}{2000}).%, \eprint{hep-th/9910133}.

\bibitem[{\citenamefont{Kotov and Strobl}(2004)}]{AT}
\bibinfo{author}{\bibfnamefont{A.}~\bibnamefont{Kotov}} \bibnamefont{and}
  \bibinfo{author}{\bibfnamefont{T.}~\bibnamefont{Strobl}},
  %(\bibinfo{year}{2004}),
   \bibinfo{note}{in preparation}.

\bibitem[{\citenamefont{Klimcik and Strobl}(2002)}]{Ctirad}
\bibinfo{author}{\bibfnamefont{C.}~\bibnamefont{Klimcik}} \bibnamefont{and}
  \bibinfo{author}{\bibfnamefont{T.}~\bibnamefont{Strobl}},
  \bibinfo{journal}{J. Geom. Phys.} \textbf{\bibinfo{volume}{43}},
  \bibinfo{pages}{341} (\bibinfo{year}{2002}).
  %, \eprint{math.sg/0104189}.

\bibitem[{\citenamefont{Ikeda}(2002)}]{IkedaCSM}
\bibinfo{author}{\bibfnamefont{N.}~\bibnamefont{Ikeda}},
  \bibinfo{journal}{JHEP} \textbf{\bibinfo{volume}{10}}, \bibinfo{pages}{076}
  (\bibinfo{year}{2002}).%, \eprint{hep-th/0209042}.

\bibitem[{\citenamefont{Roytenberg}(2002)}]{DimaCourant}
\bibinfo{author}{\bibfnamefont{D.}~\bibnamefont{Roytenberg}},
%  (\bibinfo{year}{2002}),
\eprint{math.sg/0203110}.

\bibitem[{\citenamefont{Alexandrov et~al.}(1997)\citenamefont{Alexandrov,
  Kontsevich, Schwartz, and Zaboronsky}}]{AKSZ}
\bibinfo{author}{\bibfnamefont{M.}~\bibnamefont{Alexandrov}},
  \bibinfo{author}{\bibfnamefont{M.}~\bibnamefont{Kontsevich}},
  \bibinfo{author}{\bibfnamefont{A.}~\bibnamefont{Schwartz}}, \bibnamefont{and}
  \bibinfo{author}{\bibfnamefont{O.}~\bibnamefont{Zaboronsky}},
  \bibinfo{journal}{Int. J. Mod. Phys.} \textbf{\bibinfo{volume}{A12}},
  \bibinfo{pages}{1405} (\bibinfo{year}{1997}).%, \eprint{hep-th/9502010}.

\bibitem[{\citenamefont{Alekseev and Strobl}(2004)}]{AS}
\bibinfo{author}{\bibfnamefont{A.}~\bibnamefont{Alekseev}} \bibnamefont{and}
  \bibinfo{author}{\bibfnamefont{T.}~\bibnamefont{Strobl}},
  %(\bibinfo{year}{2004}),
  \eprint{hep-th/0410183}.

\bibitem[{\citenamefont{Kontsevich}(2003)}]{Kontsevich}
\bibinfo{author}{\bibfnamefont{M.}~\bibnamefont{Kontsevich}},
  \bibinfo{journal}{Lett. Math. Phys.} \textbf{\bibinfo{volume}{66}},
  \bibinfo{pages}{157} (\bibinfo{year}{2003}).%, \eprint{q-alg/9709040}.

\bibitem[{\citenamefont{Strobl}(2004)}]{CMP1}
\bibinfo{author}{\bibfnamefont{T.}~\bibnamefont{Strobl}},
  \bibinfo{journal}{Commun. Math. Phys.} \textbf{\bibinfo{volume}{246}},
  \bibinfo{pages}{475} (\bibinfo{year}{2004}).%, \eprint{hep-th/0310168}.

\bibitem[{\citenamefont{Crainic and Fernandes}(2003)}]{CFm1}
\bibinfo{author}{\bibfnamefont{M.}~\bibnamefont{Crainic}} \bibnamefont{and}
  \bibinfo{author}{\bibfnamefont{R.~L.} \bibnamefont{Fernandes}},
  \bibinfo{journal}{Annals of Mathematics} \textbf{\bibinfo{volume}{157}},
  \bibinfo{pages}{575} (\bibinfo{year}{2003}).%, \eprint{arXiv:math.DG/0105033}.

\bibitem[{\citenamefont{Strobl}(1999)}]{TSsugra}
\bibinfo{author}{\bibfnamefont{T.}~\bibnamefont{Strobl}},
  \bibinfo{journal}{Phys. Lett.} \textbf{\bibinfo{volume}{B460}},
  \bibinfo{pages}{87} (\bibinfo{year}{1999}).
  %, \eprint[http://arXiv.org/abs]{hep-th/9906230}.

\bibitem[{\citenamefont{Bojowald et~al.}(2004)\citenamefont{Bojowald, Kotov,
  and Strobl}}]{CMP-BKS}
\bibinfo{author}{\bibfnamefont{M.}~\bibnamefont{Bojowald}},
  \bibinfo{author}{\bibfnamefont{A.}~\bibnamefont{Kotov}}, %\bibnamefont{and}
  \bibinfo{author}{\bibfnamefont{T.}~\bibnamefont{Strobl}},
  \eprint{math.DG/0406445}, \bibinfo{note}{to be published in 
  J. Geom. Phys.} .
%  (\bibinfo{year}{2004}), \bibinfo{note}{submitted to Commun. Math. Phys.}

\end{thebibliography}

\end{document}